# The elephant in the room: multi-authorship and the assessment of individual researchers

George A. Lozano

Estonian Centre of Evolutionary Ecology, 15 Tähe Street, Tartu, Estonia, 50108

*dr.george.lozano@gmail.com*

**Abstract**

When a group of individuals creates something, credit is usually divided among them. Oddly, that does not apply to scientific papers. The most commonly used performance measure for individual researchers is the h-index, which does not correct for multiple authors. Each author claims full credit for each paper and each ensuing citation. This mismeasure of achievement is fuelling a flagrant increase in multi-authorship. Several alternatives to the h-index have been devised, and one of them, the individual h-index ($h_I$), is logical, intuitive and easily calculated. Correcting for multi-authorship would end gratuitous authorship and allow proper attribution and unbiased comparisons.

The pages of most bibliometric, information science and research policy journals are replete with discussions and proposals of fair and unbiased methods to evaluate the performance of individual researchers. Some address the best methods to compare researchers in different fields [1, 2]. Others emphasize the cost effectiveness of research or the impact the money spent on a research project [3-5]. Some compare researchers of different "age" or years of activity in a given field [6, 7]. Others incorporate not only citations in peer reviewed journals but also presence and traffic on the internet [8, 9]. As useful as these proposals and methods might be, it seems that except for a few lone voices, the academic community has tacitly agreed to ignore one of the most important factors affecting all of these evaluation methods: multi-authorship.

Over the past 50 years the number of authors per paper in science has been steadily increasing [10-13]. Single authorship used to be the norm, but these days it is extremely uncommon. The increase in the number of authors per paper is caused by many other factors, independently of the increasingly interdisciplinary nature of science [14]. In many fields, including my own, studies of similar difficulty and complexity that even 20 years ago would have had 2 or 3 authors now usually have 5 to 10 authors.

Many authors have addressed this rise in multi-authorship. Several have expressed dismay at the proliferation of "honorary", "gift" and "guest" authorship [15, 16]. Most have questioned the current meaning of authorship; with so many "authors", credit, accountability and responsibility cannot be the same as before [17]. Some have tried to glean some information from the order in which authors are listed [18], and have even suggested alternative ways of citing papers [19]. Others have argued that given the currently waning authorship standards, authorship could and should be extended reviewers [20] and "editing services" [21]. Some have documented that in some fields acceptance and citation rates [22] are higher for multi-authored papers. However, few have addressed the fact that in the current system, regardless of the number of authors, each author can claim full credit for each paper and each citation [23].

Many measures of scientific achievement are available. Previously, we used to consider mostly the number of papers published and the purported quality of the journals where the papers were published. So, we used to consider productivity and reputation, but impact only indirectly. However, in the last few years the general scientific community has coalesced around an index that integrates productivity (numbers of papers published) and impact (citation rate of these papers) into a single number: the h-index [24]. An author's h-index is the number (n) of publications that have $\geq$ n citations. Papers with $\geq$ n citations are referred as being part of the "h-core".

Several modifications of the h-index have been derived to emphasize different types of accomplishments and/or favour different biases [1]. However, most of these indices are not necessary if the h-index is used only for what it was originally intended. The h-index was meant to be used to compare researchers in the same field and at the same stage of their careers. Functionally, that "same stage" can be defined not by chronological or professional age, but by whether the researchers happen to be applying for the same job, or advancing to the same level (e.g., tenure), in a given field, at similar institutions, and at about the same time. Timing is important because citation patterns have changed considerably even in the past 20 years [25]. Indices that consider other details about citation distributions might be useful for large scale bibliometric analyses, but probably unnecessary when examining the citation patterns of only a handful of applicants. Although these other indices are extensively discussed and analysed in the bibliometric literature, they are yet to be widely adopted by the general scientific community, which seems to favour the h-index.

The h-index is included as standard information for individual researchers by the most widely used scholarly search engines. Google Scholar includes the h-index, a so-called i10-index, and a "previous 5 years" h-index. The i10-index is the number of papers with 10 citations of more. The "previous 5 years" h-index only includes citations in the past 5 years. The numbers 10 and 5 are chosen completely arbitrarily, merely because of our pentadactyl and bipedal ancestry. In addition to the h-index, Thomson Reuter's Web of Knowledge provides the total citations, the number of citing articles, both also presented excluding self-citations, and along with the citations per year and citations per paper, the latter two uncorrected for self-citations. Elsevier's Scopus provides the h-index, but only considering papers published after 1995, along with the total number of citations and co-authors. Oddly, citations for pre-1995 papers are listed with each paper, but they are not included in the h-index calculation. Hence, all of these popular search engines include the h-index and provide several other indices of "individual" performance, but none of them are corrected for multiple authorship.

A problem with evaluating individual researchers using the h-index, a problem openly acknowledged by Hirsch in the original paper [24], is that the h-index does not account for the number of authors in a paper. A citation of a paper with one author counts as **one** citation, and a citation of a paper with 7 authors counts as one citation for each of the 7 authors, or **seven** citations. There is no cost to adding more authors, gratuitously or deservedly. In fact, when more people are included, more credit is automatically created because more people can claim full ownership of the paper. By magically multiplying papers and citations by the number of authors, the current system is fuelling the increase in multi-authorship and its associated problems.

To address this problem, several variants of the h-index have already been developed. For instance, Batista *et al.* [2] suggested dividing the raw h-index by the average number of authors of the papers in the raw h-core. They called this index the individual h-index ($h_i$). Schreiber [26] suggested counting papers in the h-core fractionally, dividing them by the number of authors, which "yields an effective number which is utilized to define the $h_m$-index as that effective number of papers that have been cited $h_m$ or more times." This sounds a little complicated, and it might explain why this index has not been more widely adopted. In simpler terms, it requires calculating the raw h-index first and then, instead of counting every paper in the h-core as "one", counting a paper with 2 authors as 0.5, one with 3 authors as 0.333, etc. A problem with both of these indices is that they still require calculating the raw h-index first, and the inclusion of papers in the h-core still ignores multiple authorship. A second problem is that, unlike the original h-index, these methods require the use of fractions, and it seems people just do not like fractions, even academicians. Finally, a peculiar and perhaps undesirable effect of the $h_i$ is that one additional citation of a heavily multi-authored paper could elevate this paper into the h-core, and in doing so, actually **decrease** the author's $h_i$.

A third, more intuitive and easily derived method is to first, for each paper, divide the number of citations by the number of authors, then round that number down to the nearest whole number, and finally place the papers in order of citations per author [27]. This index, also referred to as the individual h-index ($h_I$), is defined as the number of papers (n) with $\geq$ n **citations per author.** The $h_I$ uses the same logic as the raw h-index, but eliminates the multi-authorship problem before determining the h-core. This method avoids the problem of new heavily multi-authored papers decreasing an author's h-index, eliminates the multi-authorship issue before determining the h-core, spreads the credit (citations) evenly among the contributors, gives a more accurate estimate of the per-author impact, and discourages gratuitous co-authorship. Additionally, the (h$_I$) is easy to determine, almost as easy as calculating the raw h-index, and it only requires the use of whole numbers. Finally, unlike the raw h-index, which can be easily manipulated by gratuitous self-citations [28, 29], pushing a given paper up into the "h$_I$ core" requires at least as many citations as there are authors in the paper.

To account for multi-authorship in personal evaluations, it has been suggested that only a certain reasonable maximum number of publications ought to be considered [30]. Several funding agencies already do that, but they still request complete publication records and they still never account for multi-authorship. It has also been suggested that authorship should be replaced by "contributorship", whereby all contributors and their individual contributions are listed on the paper, much like credits at the end of a movie [31]. Many authors and job and grant applicants are required to do that now, but it is unclear exactly how this information is being used. Some journals include this information in the paper, but even then, only in addition to the traditional authors' list. In any case, the same problem remains: there is nothing preventing all authors from claiming crucial participation in all aspects of the study. It is as if journal editors, grant evaluators and hiring committees have decided to fiercely tackle the multi-authorship problem by gently encouraging authors to think about it a little.

Because of its simplicity and intuitive appeal, the h-index has become the most widely accepted measure of productivity and impact. However, it does not account for multiple authors and hence it does not measure individual performance. Fortunately, simple, easily determined alternatives do exist. A paper with 2 authors ought to count as half a paper for each author, and a citation of a paper with 2 authors ought to count as half a citation for each author. When we start pro-rating papers and citations, gratuitous authorship will quickly cease and proper individual attribution and comparisons will again be possible.

ACKNOWLEDGEMENTS. I thank the University of Tartu for giving me free access to their online collections. This is contribution number 1306 of the ECEE (reg. no. 80355697).

*George A. Lozano*

*Estonian Centre of Evolutionary Ecology, 15 Tähe Street, Tartu, Estonia, 50108*

*E-mail*: [dr.george.lozano@gmail.com](mailto:dr.george.lozano@gmail.com)